\documentclass{PoS}

\title{\center Analytic Neutrino Oscillation Probabilities in Matter: Revisited\footnote{FERMILAB-CONF-17-598-T}
}

\ShortTitle{Neutrinos Propogation in Matter}

\author{\speaker{Stephen J. Parke}\thanks{I wish to thank the organizers for the fabulous Swedish hospitality!}\\
        Theoretical Physics Department, Fermi National Accelerator Laboratory,\\ %P.\ O.\ Box 500, 
        Batavia, IL 60510, USA \\ 
        E-mail: \email{parke@fnal.gov}  }

\author{Peter B.~Denton \\
        Niels Bohr International Academy, Niels Bohr Institute, University of Copenhagen,\\
         Blegdamsvej 17, 2100, 
         Copenhagen, Denmark\\
       E-mail: \email{peterbd1@gmail.com}}
       
\author{Hisakazu Minakata\\Instituto F\'{\i}sica Te\'{o}rica, UAM/CSIC, %Calle Nicol\'as Cabrera 13-15, 
Cantoblanco E-28049 Madrid, Spain \\
\& Research Center for Cosmic Neutrinos, ICRR, University of Tokyo, %Kashiwa, Chiba 277-8582, 
Japan\\
E-mail: \email{hisakazu.minakata@gmail.com} \\
}

\abstract{We summarize our recent paper on neutrino oscillation probabilities in matter, explaining the importance, relevance and need for simple, highly accurate approximations to the neutrino oscillation probabilities in matter. Simple expressions for the neutrino mixing angles and mass squared differences in matter are given in an Appendix. Using these in the vacuum oscillation probabilities instead of the vacuum values, gives an excellent approximation to the oscillation probabilities in matter.
}

\FullConference{The 19th International Workshop on Neutrinos from Accelerators-NUFACT2017\\
		25-30 September, 2017\\
		Uppsala University, Uppsala, Sweden}

\begin{document}

\section{Neutrino Propagation in Matter}

The evolution of a neutrino flavor state in matter is given by\\[1mm]
\indent $i\frac{d}{dx}   \nu =  H \nu$  with  $\nu = \left( \begin{array}{c} \nu_e \\ \nu_\mu \\ \nu_\tau \end{array} \right)$ and
%\begin{eqnarray}
$
H=\frac{1}{2E} \left\{ U \left[ \begin{array}{ccc} 0 & 0 & 0 \\ 0 & \Delta m^2_{21} & 0 \\ 0&0& \Delta m^2_{31} \end{array} \right] U^\dagger  +\left[ \begin{array}{ccc} a(x) & 0 & 0 \\ 0 & 0 & 0 \\ 0&0&0 \end{array} \right]\right\},
$  \\[1mm]
%\end{eqnarray}
where ``a'' is the matter potential, $a= 2 \sqrt{2} G_F N_e E$.

We have developed a {\it single} perturbative expansion for the oscillation probabilities  in constant matter that satisfies the following criteria, see \cite{Denton:2016wmg};
\begin{enumerate}
\item valid and accurate for all baseline divided by neutrino energy and all values of matter potential, i.e. over the full $(L/E, ~ Y\rho E)$ plane,
\item  has the universal form for the L/E dependence of the oscillation probabilities i.e. three $\sin$ squared terms and a CP violating triple $\sin$ term,
\item  since the atmospheric and solar resonance have to be dealt with in a non-perturbative fashion, we need to use $\sqrt{\cdots} ~$ function but will use nothing more complex\footnote{Compared to the exact results which involve the $\cos[\frac{1}{3} \arccos(\cdots)]$ expressions that appear in \cite{Zaglauer:1988gz}. },
\item the form of the mass eigenvalues squared in matter is particularly simple which leads to simple forms for the mixing angles in matter at zeroth order, providing an enhanced understand of oscillation probabilities in matter.
\end{enumerate}
Our perturbative expansion should be compared to other perturbative expansions in the literature, see \cite{Arafune:1997hd}, which do not satisfy {\it all} of the above criteria.

To develop a perturbation theory for the neutrino mass squared's in matter as well as the elements of MNS in matter, we first need to take care of the resonance regions non-perturbatively.
We start by splitting the Hamiltonian, $H$, into two pieces, the diagonal part, $H_0$,  and the non-diagonal part $H_1$ such that $H=H_0+H_1$:
\begin{eqnarray}
H_0  & =  & \frac{1}{2E} \left[ \begin{array}{ccc}
\lambda_a & &\\
&\lambda_b  &\\
&  &\lambda_c \\
\end{array}
\right]  
\label{eqn:H0}
 \quad {\rm where} \quad  \left\{ \begin{array}{l}  \quad \lambda_a \equiv ~~a+s^2_{13} \Delta m^2_{ee} \\
\quad  \lambda_b \equiv ~~(c^2_{12}-s^2_{12}) \Delta m^2_{21} \\ \quad \lambda_c \equiv ~~c^2_{13} \Delta m^2_{ee} \end{array}
 \right. ,
\end{eqnarray}
where $ \Delta m^2_{ee} \equiv  \Delta m^2_{31} -s^2_{12}  \Delta m^2_{21}= \Delta m^2_{32} +c^2_{12}  \Delta m^2_{21}$.
Note, $\lambda_a$, $\lambda_b$ and $\lambda_c$ are the asymptotic values\footnote{The constant $s^2_{12} \Delta m^2_{21}$ has been subtracted from all eigenvalue compared to \cite{Denton:2016wmg}.} of the mass squared eigenstates as $|a| \rightarrow \infty$, including terms of ${\rm O}(a^0)$.

 The non-diagonal part, $H_1$, is given by
\begin{eqnarray}
H_1 &  =  &  s_{13} c_{13}  \frac{ \Delta m^2_{ee} }{2E} \left[ \begin{array}{ccc}
  & & 1\\
& 0  &\\
 1 &  &   
\end{array}
\right] 
%\end{eqnarray}
%\begin{eqnarray}
%H_1 =
%\\[3mm] && \hspace*{-2cm}  
+ ~c_{13}~s_{12} c_{12}  \frac{ \Delta m^2_{21} }{2E} \left[ \begin{array}{ccc}
 &  1& \\
 1 &   &  0\\
  &  0 &  \\
\end{array}
\right] 
%\\[3mm] && \hspace*{-2cm}  
-  ~s_{13}~s_{12} c_{12}  \frac{ \Delta m^2_{21} }{2E} \left[ \begin{array}{ccc}
 &  0 & \\
 0 &   &  1\\
  &  1 &  \\
\end{array}
\right].  \quad ~~~
\end{eqnarray}
Given that 
%\newpage
%$s_{13} c_{13} \sim 0.15$, $~c_{13} s_{12} c_{12} \Delta m^2_{21}/\Delta m^2_{ee} \sim 0.015$ and $s_{13} s_{12} c_{12}\Delta m^2_{21}/\Delta m^2_{ee} \sim 0.002$ , 
\begin{eqnarray} 
s_{13} c_{13} \sim 0.15, \quad c_{13} s_{12} c_{12}( \Delta m^2_{21}/\Delta m^2_{ee}) \sim 0.015 \quad {\rm and} \quad s_{13} s_{12} c_{12} (\Delta m^2_{21}/\Delta m^2_{ee}) \sim 0.002, \nonumber 
\end{eqnarray}
there is a hierarchy in the size of the three terms in $H_1$. So it is natural to perform first a rotation in the 1-3 space followed by a rotation in the 1-2 space.  After, these rotations, the level crossings that existed in $H_0$ of eqn \ref{eqn:H0}, as one varied the matter potential, no longer exist in the new $H_0$ given by:
\begin{eqnarray}
H_0   =   \frac{1}{2E} \left[ \begin{array}{ccc}
\lambda_1 & &\\
&\lambda_2  &\\
&  &\lambda_3 \\
\end{array}
\right]  
%\end{eqnarray}
& \quad  {\rm and} \quad &
%\begin{eqnarray}
H_1   =    \sin (\phi-\theta_{13}) s_{12} c_{12}  \frac{ \Delta m^2_{21} }{2E} \left[ \begin{array}{ccc}
 0 &0 & -s_\psi\\
0& 0  & c_\psi\\
 -s_\psi & c_\psi & 0  \\
\end{array}
\right]  
\label{eqn:Hpert}
\end{eqnarray}
where $\lambda_i$ are the square of the neutrino mass in matter and $\phi$, $\psi$ are the mixing angles $\theta_{13}$, $\theta_{12}$ in matter.  Note, in vacuum, $ \sin (\phi-\theta_{13})=0$ so that $H_1=0$. Expressions for $\lambda_i$, $\phi$ and $\psi$ will be given in next section.

At this point, one could perform a further rotation.  For NO, if one performs an additional rotation in the 1-3 space, then the new $H_1$ will be proportional to  $ \sin (\phi-\theta_{13}) c_\psi $ whose magnitude for all ``a'' is $< s_{13}$. Similarly for IO, if the rotation is performed in the 2-3 space and  the new $H_1$ wiil be proportional to  $ \sin (\phi-\theta_{13}) s_\psi $ . These rotations would significantly improve the 0th order approximation especially in the region where  $ \sin (\phi-\theta_{13}) \sim c_{13} $.

Instead, to keep one perturbative expansion for both mass orderings, we will do perturbation theory using the results of the first two rotations, i.e. eqn \ref{eqn:Hpert}.

\section{A Simple, Accurate Method for Calculate Oscillation Probabilities in Matter}

A simple and accurate way to evaluate oscillation probabilities, see \cite{Denton:2016wmg}, is given in this section. Details as to the why's and how's of this method are contained in this paper.

After performing a rotation in  the 1-3  space: % $\theta_{13} \rightarrow \phi$:
\begin{eqnarray}
\lambda_0 = \lambda_b,  \quad% \nonumber \\[3mm]
\lambda_{\pm}  &= & \frac{1}{2} \left(\lambda_c+\lambda_a \pm {\rm sign}( \Delta m^2_{ee}) \sqrt{ (\lambda_c-\lambda_a)^2 + (2 s_{13} c_{13} \Delta m^2_{ee})^2} ~\right)
\nonumber \\[3mm]
& = &  \frac{1}{2} \left(\Delta m^2_{ee} +a ~\pm {\rm sign}( \Delta m^2_{ee}) \sqrt{ (\Delta m^2_{ee} \cos 2 \theta_{13} -a)^2 + (2 s_{13} c_{13} \Delta m^2_{ee})^2} ~\right)
\nonumber \\[3mm]
\sin \phi & = & \sqrt{ \frac{\lambda_+ -\lambda_c}{\lambda_+ -\lambda_-} }   \quad  {\rm with}~0 \leq \phi \leq \pi/2, \nonumber \\
{\rm which ~satisfies} \nonumber \\
\phi(a) & = & \frac{\pi}{2}- \phi(2 \Delta m^2_{ee} \cos \theta_{13} -a) \quad {\rm and}  \quad \phi_{NO}(a) = \phi_{IO}(-a).
%\quad  {\rm OR}  \quad \sin 2 \phi =\frac{2 s_{13} c_{13} \Delta m^2_{ee}}{\lambda_+ -\lambda_-}  
\end{eqnarray}
$\phi$ is $\theta_{13}$ in matter and $(\lambda_a,\lambda_b,\lambda_c) \rightarrow (\lambda_-,\lambda_0,\lambda_+)$.    When 
 $|\lambda_c-\lambda_a| \gg 2 s_{13} c_{13} |\Delta m^2_{ee}|$ then $(\lambda_-,\lambda_+) \approx (\lambda_a,\lambda_c) ~{\rm or} ~(\lambda_c,\lambda_a)$. If $a=0$, $\phi=\theta_{13}$ and $(\lambda_-,\lambda_0,\lambda_+) =(0,(c^2_{12}-s^2_{12})\Delta m^2_{21},\Delta m^2_{ee})$.  \\
 
It is simple to show that 
\begin{eqnarray} 
\sin(\phi-\theta_{13}) ~~= ~~{\rm sign}(a \Delta m^2_{ee}) \sqrt{(\lambda_- - a c^2_\phi )/\Delta m^2_{ee}} ~~\approx ~~s_{13} c_{13} (a/\Delta m^2_{ee}) + {\cal O}[(a/\Delta m^2_{ee})^2],  \nonumber
\end{eqnarray}
 which ultimately will determine the size of the perturbing Hamiltonian.  Note, that for a neutrino energy of 3 GeV and earth crust density $\sin(\phi-\theta_{13})  \approx 0.04$. \\

%$P(\Delta \lambda_{jk}, \phi, \theta_{12}, \theta_{23}, \delta)$    with $j,k=(-,0,+)$

Then, performing a rotation in  the 1-2 space: with  $\lambda_3 = \lambda_+$ and
\begin{eqnarray}
% \nonumber \\[3mm]
\lambda_{2,1} &  = & \frac{1}{2} \left(\lambda_0+\lambda_- \pm \sqrt{ (\lambda_0-\lambda_-)^2 + (2 \cos(\phi-\theta_{13}) s_{12} c_{12} \Delta m^2_{21})^2} ~\right) 
\nonumber \\[3mm]
& \approx &  \frac{1}{2} \left(\Delta m^2_{21} \cos 2\theta_{12} + a~ c^2_{13}  ~\pm \sqrt{ (\Delta m^2_{21} \cos 2\theta_{12} - a ~c^2_{13})^2 + (2 s_{12} c_{12} \Delta m^2_{21})^2} ~\right) \nonumber \\ & & \hspace*{9cm} {\rm for} ~|a/\Delta m^2_{ee}| \ll 1
\nonumber \\[1mm]
\sin \psi & = & \sqrt{ \frac{\lambda_2 -\lambda_0}{\lambda_2 -\lambda_1} } ,  \quad  {\rm with}~0 \leq \psi \leq \pi/2.  \nonumber \\
{\rm which ~satisfies} \nonumber \\
\psi(a) & \approx & \frac{\pi}{2}- \psi(2 \Delta m^2_{21} \cos \theta_{12}/c^2_{13} -a) \quad {\rm and}  
\quad \phi_{NO}(a) \approx \phi_{IO}(a).
% \quad  {\rm OR}  \quad \sin 2 \psi =\frac{2\cos(\phi-\theta_{13}) s_{12} c_{12} \Delta m^2_{21}}{\lambda_2 -\lambda_1}  
\end{eqnarray}
%It is worth noting that  $\sin(\phi-\theta_{13}) = {\rm sign}(a \Delta m^2_{ee}) \sqrt{(\lambda_+ - a s^2_{13} - \Delta m^2_{ee})/(\lambda_{+}-\lambda_{-})}$.
 $\psi$ is $\theta_{12}$ in matter and $(\lambda_-,\lambda_0,\lambda_+) \rightarrow (\lambda_1,\lambda_2,\lambda_3)$.
When 
 $|\lambda_0-\lambda_-| \gg 2 s_{12} c_{12} \Delta m^2_{21}$ then $(\lambda_1,\lambda_2) \approx (\lambda_-,\lambda_0) ~{\rm or} ~(\lambda_0,\lambda_-)$. 
 If $a=0$, $\psi=\theta_{12}$ and $(\lambda_1,\lambda_2,\lambda_3) =(-s^2_{12} \Delta m^2_{21},c^2_{21} \Delta m^2_{21}, \Delta m^2_{ee})$, thus, in vacuum,  $\Delta \lambda_{jk} \equiv \lambda_j-\lambda_k = \Delta m^2_{jk}$. See Appendix for further details.
%In the Appendix, we give approximate expressions for the $\lambda$'s in all regions of interest for NO.\\
 
% \newpage 
To calculate the oscillation probabilities, to 0th order,  use the above $\Delta \lambda_{jk} $ instead of $\Delta m^2_{jk}$ and replace the vacuum MNS matrix as follows\footnote{For the rest of this paper we use the standard parameterization of the MNS matrix for the reader's convenience, as oppose to the parametrization used in \cite{Denton:2016wmg}.}
\begin{eqnarray}
U^0_{MNS} \equiv U_{23}(\theta_{23})U_{13}(\theta_{13},\delta)U_{12}(\theta_{12})  
& \Rightarrow  & U^M_{MNS} \equiv U_{23}(\theta_{23})U_{13}(\phi,\delta)U_{12}(\psi).
\end{eqnarray}
That is, replace
\begin{eqnarray}
\Delta m^2_{jk}   \rightarrow  \Delta \lambda_{jk} \quad 
& \theta_{13}   \rightarrow  \phi, \quad 
\theta_{12}   \rightarrow & \psi,
%U_{PMNS}(\theta_{13},\theta_{12},\theta_{23},\delta)  & \rightarrow & U_{PMNS}(\phi,\psi,\theta_{23},\delta) 
\end{eqnarray}
it is that simple. $\theta_{23}$ and $\delta$ remain unchanged.  Our expansion parameter is $\left| \sin(\phi-\theta_{13}) ~s_{12} c_{12} ~\frac{\Delta m^2_{21}}{\Delta m^2_{ee}} \right| \leq 0.015 $, which is small and vanishes in vacuum, so that our perturbation theory reproduces the vacuum oscillation probabilities exactly.
A summary of the relevant expressions are given in Fig. 1 and alternative summary using a more conventional notation is given in Appendix II.

\subsection{Higher Orders}
If the 0th order is not accurate enough, going to 1st order is simple and gives another two orders of magnitude in accuracy.  First the eigenvalues $\lambda_j$ remain unchanged but the mixing matrix is modified by
\begin{eqnarray}
U^M_{MNS} & \Rightarrow & V \equiv U^M_{MNS} (1+W_1),
\end{eqnarray}
where the matrix $W_1$ is given by\footnote{The phase in $W_1$ differ from \cite{Denton:2016wmg} because here we use the standard parameterization of $U_{MNS}$.}
\begin{eqnarray}
W_1 = \sin(\phi-\theta_{13}) ~s_{12} c_{12} ~\Delta m^2_{21} \left( \begin{array}{ccc}
 0 & 0 & -s_\psi e^{-i\delta} / \Delta \lambda_{31} \\
 0 &0 & +c_\psi e^{-i\delta} / \Delta \lambda_{32} \\
 +s_\psi e^{+i\delta}  / \Delta \lambda_{31} \quad  & -c_\psi e^{+i\delta} / \Delta \lambda_{32} \quad &  0
 \end{array} \right).  %\nonumber
 \end{eqnarray}
 The $\Delta \lambda_{jk}$ and the $V$-mixing matrix can be used to calculate the oscillation probabilities and improve the accuracy so that 
 $\Delta P < 10^{-6}$.  The next highest order is also discussed in \cite{Denton:2016wmg}.

\begin{figure}[t]
\begin{center}
     \includegraphics[width=.9\textwidth]{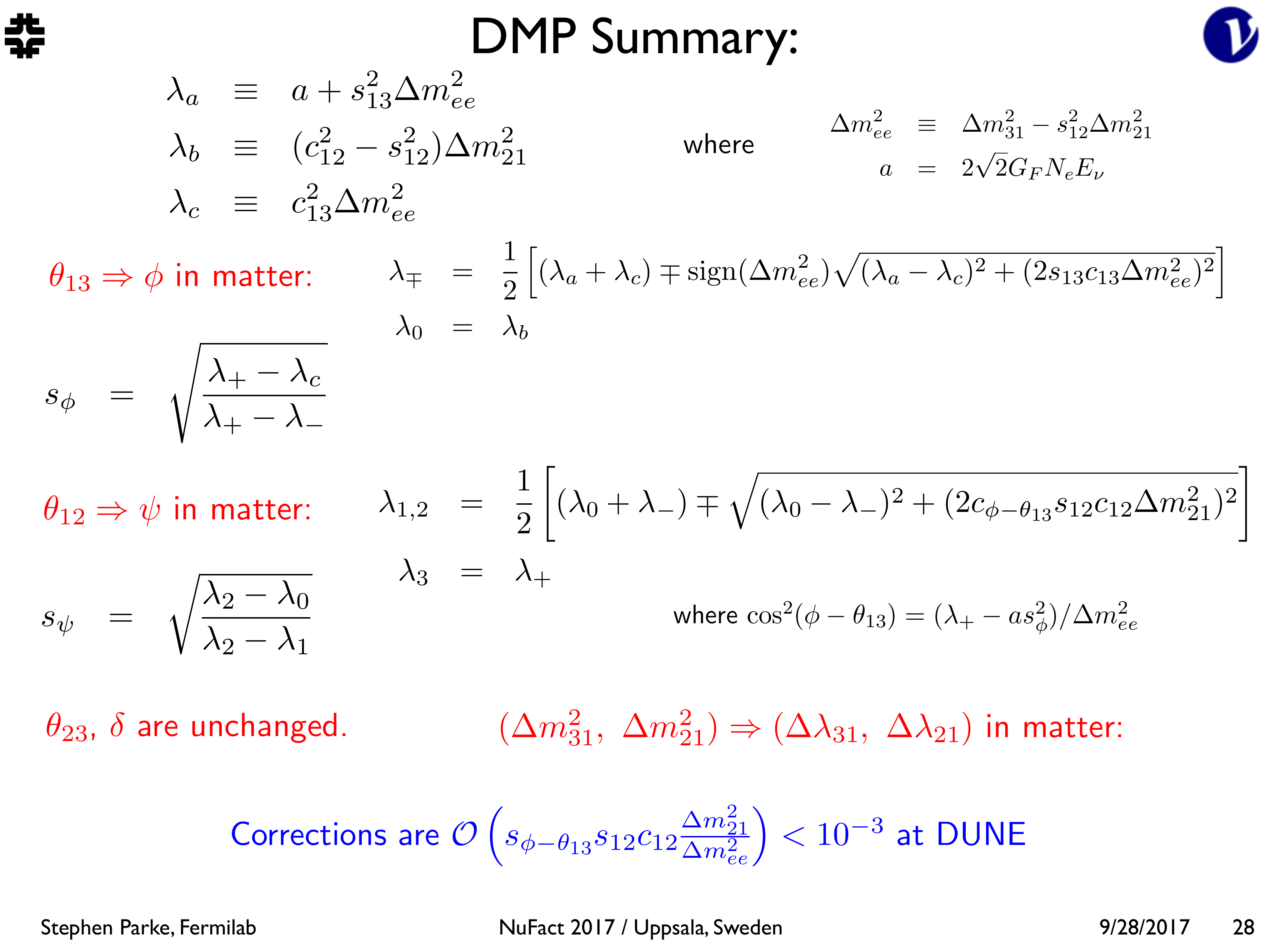}
     \caption{Summary of DMP perturbation theory, 0th order, see \cite{Denton:2016wmg}, for the mixing angles and mass squared eigenvalues in matter.  Replacing $(\theta_{12}, ~\theta_{13}, ~\theta_{23}, ~\delta, ~\Delta m^2_{31}, ~\Delta m^2_{21})$ with  $(\psi, ~\phi, ~\theta_{23}, ~\delta, ~\Delta \lambda_{31}, ~\Delta \lambda_{21})$ in the vacuum oscillation expressions gives oscillation probability in matter to 0th order. Only 6 square root operations are required to go from the vacuum to matter parameters, not counting the simple $(+,-,*,/)$ operations. No other time consuming operations like sine, cosine, arcsine, arccosine etc are needed.}
     \end{center}
     \label{fig:dmp}
     \end{figure}

\begin{figure}[b]
     \includegraphics[width=.49\textwidth]{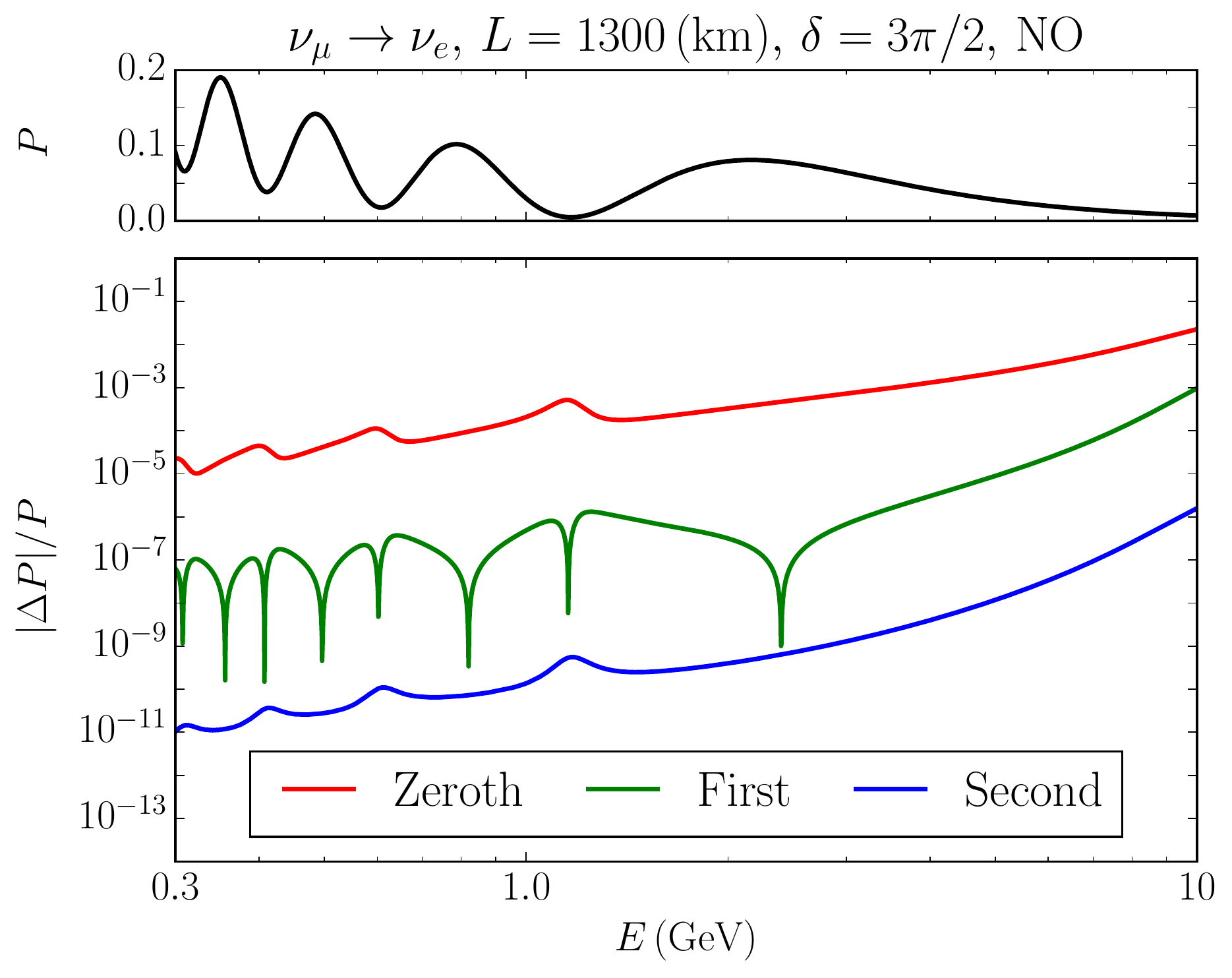}
 \includegraphics[width=.49\textwidth]{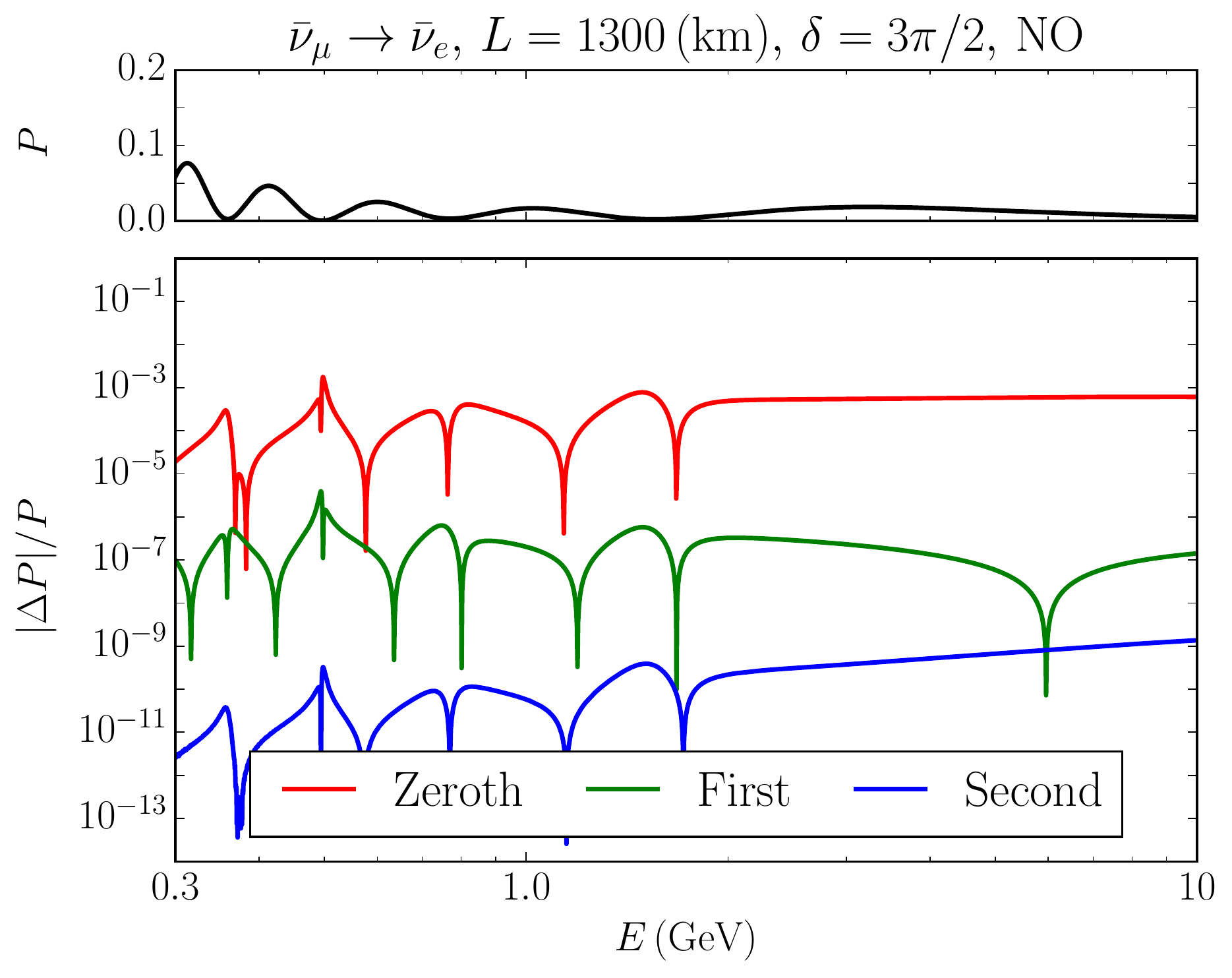}
     \caption{Dune $\nu_\mu \rightarrow \nu_e$ and $\bar{\nu}_\mu \rightarrow \bar{\nu}_e$ appearance probabilities, top panel. The bottom panel, shows the fractional difference between the 0th, 1st and 2nd order approximations to the exact probabilities assuming a constant matter density. }
     \label{fig:mu2e}
     \end{figure}

\clearpage
In Fig. \ref{fig:mu2e} we have compared the exact oscillation probability with our approximation. One sees that the 0th order oscillation probabilities, relevant for the DUNE experiment, have a difference, from the exact calculation, $|\Delta P| <  10^{-4}$ and $|\Delta P|/P <  10^{-3}$. Higher orders are even more accurate.

\section{Conclusions}

We have summarized our perturbation theory in matter for the neutrino oscillation probabilities, that gives the neutrino mass eigenvalues squared in matter in simple terms only involving the $\sqrt{\cdots}~$ function. We also show how the mixing angles in matter can be obtained directly once one knows the 
matter mass squareds.   Higher orders are simple to obtain and increase the accuracy by about two orders of magnitude per order. However, the zeroth order approximations are good enough for all current and future accelerator experiments; T2K, NO$\nu$A, DUNE and T2HK/T2HKK due to the uncertainties associated with the matter density profile, height and shape, between neutrino production and detection.

\section{Appendix}

For NO we give approximate expressions for the  $\lambda$'s, in the different regions of interest;
%\begin{eqnarray}
%\lambda_3 & \approx & \left\{ \begin{array}{ll}
%a+s^2_{13} \Delta m^2_{ee},  & \quad (a-\Delta m^2_{ee})/\Delta m^2_{ee} \gg 1  \nonumber \\
%\frac{1}{2} \left(\Delta m^2_{ee} +a  \right. &  \nonumber \\
%\quad \quad \left. + \sqrt{ (\Delta m^2_{ee} \cos 2 \theta_{13} -a)^2 + (2 s_{13} c_{13} \Delta m^2_{ee})^2} ~\right), &  \quad   a/\Delta m^2_{ee} \approx 1 \nonumber \\
%\Delta m^2_{ee} +s^2_{13} a, &  \quad |a| /\Delta m^2_{ee} \ll 1  \nonumber \\
%c^2_{13} \Delta m^2_{ee}, & \quad  (\Delta m^2_{ee}-a)/\Delta m^2_{ee} \gg 1.  \nonumber 
%\end{array}
%\right.
%\end{eqnarray}
\begin{eqnarray}
\lambda_3 & \approx & \left\{ \begin{array}{ll}
a+s^2_{13} \Delta m^2_{ee},  & \quad a \gg  \Delta m^2_{ee}  \nonumber \\
\frac{1}{2} \left(\Delta m^2_{ee} +a  \right. &  \nonumber \\
\quad \quad \left. + \sqrt{ (\Delta m^2_{ee} \cos 2 \theta_{13} -a)^2 + (2 s_{13} c_{13} \Delta m^2_{ee})^2} ~\right), &  \quad   a \approx \Delta m^2_{ee}  \nonumber \\
\Delta m^2_{ee} +s^2_{13} a, &  \quad |a| \ll \Delta m^2_{ee}   \nonumber \\
c^2_{13} \Delta m^2_{ee}, & \quad  -a \gg \Delta m^2_{ee}.  \nonumber 
\end{array}
\right.
\end{eqnarray}

One can obtain similar expressions for $\lambda_-$ using $\lambda_- = \Delta m^2_{ee} + a -\lambda_3$.
%\begin{eqnarray}
%\lambda_- & = & \left\{ \begin{array}{ll}
%c^2_{13} \Delta m^2_{ee},  & \quad (a-\Delta m^2_{ee})/\Delta m^2_{ee} \gg 1  \nonumber \\
%\frac{1}{2} \left(\Delta m^2_{ee} +a  \right. &  \nonumber \\
%\quad \quad \left. - \sqrt{ (\Delta m^2_{ee} \cos 2 \theta_{13} -a)^2 + (2 s_{13} c_{13} \Delta m^2_{ee})^2} ~\right), &  \quad   a/\Delta m^2_{ee} \approx 1 \nonumber \\
%c^2_{13} a, &  \quad |a| /\Delta m^2_{ee} \ll 1  \nonumber \\
 %a+s^2_{13} \Delta m^2_{ee},& \quad  (\Delta m^2_{ee}-a)/\Delta m^2_{ee} \gg 1.  \nonumber 
%\end{array}
%\right.
%\end{eqnarray}
Also,
\begin{eqnarray}
\lambda_1 & \approx & \left\{ \begin{array}{ll}
\cos 2\theta_{12} \Delta m^2_{21},  & \quad a \gg \Delta m^2_{21}  \nonumber \\
 \frac{1}{2} \left(\Delta m^2_{21}\cos 2\theta_{12} + a~ c^2_{13}   \right. &  \nonumber \\
\quad \quad \left. -\sqrt{ (\Delta m^2_{21} \cos 2\theta_{12} - a ~c^2_{13})^2 + (2 s_{12} c_{12} \Delta m^2_{21})^2} ~\right), &  \quad   a \approx \Delta m^2_{21}  \nonumber  \\
 -s^2_{12} \Delta m^2_{21}+ c^2_{12} c^2_{13} a , &  \quad |a| \ll \Delta m^2_{21}  \nonumber \\
 a+s^2_{13} \Delta m^2_{ee},& \quad  -a \gg \Delta m^2_{ee} .  \nonumber 
\end{array}
\right.
\end{eqnarray}
To obtain $\lambda_2$ in all regions of interest, we use $\lambda_2 =  \Delta m^2_{ee}+\cos 2\theta_{12} \Delta m^2_{21} + a -\lambda_3 -\lambda_1$; 
\begin{eqnarray}
\lambda_2 & \approx & \left\{ \begin{array}{ll}
c^2_{13} \Delta m^2_{ee},  & \quad a \gg \Delta m^2_{ee} \nonumber \\
\frac{1}{2} \left(\Delta m^2_{ee} +a  \right. &  \nonumber \\
\quad \quad \left. - \sqrt{ (\Delta m^2_{ee} \cos 2 \theta_{13} -a)^2 + (2 s_{13} c_{13} \Delta m^2_{ee})^2} ~\right), &  \quad   a \approx \Delta m^2_{ee}  \nonumber \\
 \frac{1}{2} \left(\Delta m^2_{21} \cos 2\theta_{12} + a~ c^2_{13}   \right. &  \nonumber \\
\quad \quad \left. +\sqrt{ (\Delta m^2_{21} \cos 2\theta_{12} - a ~c^2_{13})^2 + (2 s_{12} c_{12} \Delta m^2_{21})^2} ~\right), &  \quad   a \approx \Delta m^2_{21}  \nonumber  \\
 c^2_{12} \Delta m^2_{21}+ s^2_{12} c^2_{13} a , &  \quad |a| \ll \Delta m^2_{21}   \nonumber \\
 (c^2_{12}-s^2_{12}) \Delta m^2_{21},  & \quad  -a \gg \Delta m^2_{ee} .  \nonumber 
\end{array}
\right.
\end{eqnarray}

 Since $\psi$ and $\phi$ are given in terms of the eigenvalues earlier, approximations for all variables can be easily derived from these $\lambda$'s.  For IO, one can give similar expressions for $\lambda$'s. In Appendix III we give the eigenvalues in terms of the mixing angles.
 
 \newpage
 \section{Appendix II}
 We have written everything in terms of the matter eigenvalues, the $\lambda$'s, since if you know the  $\lambda$'s you can easily calculate the matter mixing angles, $\phi$  ($=\widetilde{\theta}_{13}$ )  and $\psi$ ($=\widetilde{\theta}_{12}$), this is summarized in Fig. 1.
   
However, the mixing angles in matter, which we denote by a $\widetilde{\theta}_{13}$ and $\widetilde{\theta}_{12}$ here, can also be calculated in the following way, using 
%the matter potential $a  = 2 \sqrt{2} G_F N_e E$ and  
$ \Delta m^2_{ee} \equiv \cos^2 \theta_{12} \Delta m^2_{31} + \sin^2 \theta_{12} \Delta m^2_{32}$,  as follows\footnote{Vacuum values to be used in calculating $\Delta m^2_{ee}$.}:
\begin{eqnarray}
\cos 2 \widetilde{\theta}_{13} & = & \frac{ (\cos 2\theta_{13} -a/\Delta m^2_{ee}) } 
{  \sqrt{(\cos 2\theta_{13}-a/\Delta m^2_{ee})^2 +  \sin^22\theta_{13}  ~ }}, 
 \label{eq:th13}  
 \end{eqnarray}
 where  $~~a  \equiv   2 \sqrt{2} G_F N_e E~~$  is the standard matter potential, and
\begin{eqnarray}
 \cos 2 \widetilde{\theta}_{12} & = &  \frac{ ( \cos 2\theta_{12} 
 - a^{\,\prime}  /\Delta m^2_{21} ) } {  \sqrt{(\cos 2\theta_{12} 
 -a^{\,\prime} /\Delta m^2_{21})^2 ~+~
  \sin^2 2 \theta_{12} \cos^2( \widetilde{\theta}_{13}-\theta_{13})~~}  }, \label{eq:th12} 
  \end{eqnarray}
where $~~a^{\,\prime}  \equiv   a \, \cos^2 \widetilde{\theta}_{13} +\Delta m^2_{ee} \sin^2  ( \widetilde{\theta}_{13}-\theta_{13} ) ~~$ is the $\theta_{13}$-modified matter potential for the 1-2 sector.
In these two flavor rotations, both $\widetilde{\theta}_{13}$ and  $\widetilde{\theta}_{12}$ are in range $[0,\pi/2]$.\\
%(The trig id  $\sin^2 \theta = (1-\cos 2\theta)/2$ is useful for evaluating $\sin \theta$ and  $\cos\theta$.)   \\

$\theta_{23}$ and $\delta$ are unchanged in matter for this approximation.\\

The neutrino mass squared differences in matter, i.e. the $\Delta m^2_{jk}$ in matter, which we denote by $\Delta \, \widetilde{m^2}_{jk}$, are given by 
   \begin{eqnarray}
  \Delta\, \widetilde{m^2}_{21}  & = & \Delta m^2_{21} \, \sqrt{(\cos 2\theta_{12} 
 - a^{\,\prime} /\Delta m^2_{21})^2 ~+~
  \sin^2 2 \theta_{12} \cos^2(\widetilde{\theta}_{13}-\theta_{13})~~} , \nonumber   \\[1mm]
   \Delta\,  \widetilde{m^2}_{31}   &=& \Delta m^2_{31} +\frac{1}{2}\,\left( ~2\, a- 3\,a^\prime  
 + \, \Delta \widetilde{m^2}_{21}  -\Delta m^2_{21}  ~\right) , \label{eq:dmsqa}   \\[1mm]
   \Delta\,  \widetilde{m^2}_{32}  & = &  \Delta \,  \widetilde{m^2}_{31} -\Delta\,   \widetilde{m^2}_{21}= \Delta m^2_{32} +\frac{1}{2}\,\left( ~2\, a- 3\,a^{\,\prime}  
 - \, \Delta \, \widetilde{m^2}_{21}  +\Delta m^2_{21}  ~\right). \nonumber
   %
   %
%  \Delta \lambda_{32}  & = &\frac{1}{4} \left[ ~ a~+ ~\Delta m^2_{ee} + 3 \,\Delta m^2_{ee}  \sqrt{(\cos 2\theta_{13}-a/\Delta m^2_{ee})^2 +  \sin^22\theta_{13} }   \right.   \nonumber \\
% & &  \left.  - 2 \,\Delta m^2_{21} \left(  \cos 2\theta_{12}  
% +\sqrt{(\cos 2\theta_{12}  - a_{12} /\Delta m^2_{21})^2 ~+~
%  \sin^2 2 \theta_{12} \cos^2(\theta^\prime_{13}-\theta_{13})~~}   \right)  ~\right] .  \nonumber  
 \end{eqnarray}
Note that the same square root\footnote{If $a=0$, then $ \widetilde{\theta}_{13}=  \theta_{13}$ and since $a^\prime=0$ then $ \widetilde{\theta}_{12}=  \theta_{12}$ and both $\sqrt{\cdots}=1$, also $\Delta \, \widetilde{m^2}_{jk}=\Delta m^2_{jk}$ for all $(j,k)$ as required. The identity $s^2_\theta=(1-\cos 2 \theta)/2$ is useful for calculating both $s_\theta$ and $c_\theta$. } 
appears  in both $ \Delta \, \widetilde{m^2}_{21}$ and $ \widetilde{\theta}_{12}$.
  To see that the $\Delta\,  \widetilde{m^2}_{31}$ and  $\Delta\,  \widetilde{m^2}_{32}$ have the right asymptotic forms, use 
 the fact that $ (\Delta \, \widetilde{m^2}_{21}  -\Delta m^2_{21}) = |a^{\, \prime} | +{\cal O}(\Delta m^2_{21})$, for $|a| \gg \Delta m^2_{21}$.\\
 
 These expressions are valid for both NO, $\Delta m^2_{31}>0$ and IO,
$\Delta m^2_{31}<0$.  For anti-neutrinos, just change the sign of $a$ and $\delta$.\\

If $P_{\nu_\alpha \rightarrow \nu_\beta}( \Delta m^2_{31}, \Delta m^2_{21}, \theta_{13}, \theta_{12},\theta_{23},\delta)$ is the oscillation probability in vacuum
 then  \\[1mm]
$P_{\nu_\alpha \rightarrow \nu_\beta}( \Delta \, \widetilde{m^2}_{31}, \Delta \, \widetilde{m^2}_{21}, \widetilde{\theta}_{13}, \widetilde{\theta}_{12},\theta_{23},\delta)$ is the oscillation probability in matter, i.e. use the same function but replace the mass squared differences and mixing angles with the matter values given in eq. \ref{eq:th13}-\ref{eq:dmsqa}. The resulting oscillation probabilities are identical%\footnote{Eqn 5.1, 5.2 and 5.3 in this Appendix are identical to eq. 2.3.5, 2.4.9 and 2.4.5, respectively of Denton et al.} 
to the zeroth order approximation given in  Denton, Minakata and Parke, \cite{Denton:2016wmg}.
%($\theta_{23}$ and $\delta$ are unchanged in matter in this approximation.)

\section{Appendix III}
One can also write the mass eigenvalues in matter, purely in terms of the mixing angles as follows:\begin{eqnarray}
\lambda_1  &=& a\, c^2_\phi c^2_\psi  + \Delta m^2_{31}\, s^2_{(\phi-\theta_{13})} c^2_\psi +\Delta m^2_{21} \,(s_\psi c_{12} - c_{(\phi-\theta_{13})} c_\psi s_{12})^2 -  s^2_{12}  \Delta m^2_{21} \nonumber \\
 \lambda_2 &=& a\, c^2_\phi s^2_\psi  + \Delta m^2_{31}\, s^2_{(\phi-\theta_{13})} s^2_\psi +\Delta m^2_{21} \,(c_\psi c_{12} + c_{(\phi-\theta_{13})} s_\psi s_{12})^2 -  s^2_{12}  \Delta m^2_{21}  \nonumber  \\
\lambda_3 &=& a\, s^2_\phi ~~  +  ~~\Delta m^2_{ee}\, c^2_{(\phi-\theta_{13})} =\lambda_+  \\
%\lambda_3 &=& a\, s^2_\phi ~~  +  ~~\Delta m^2_{31}\, c^2_{(\phi-\theta_{13})} ~~  + ~~ \Delta m^2_{21} \,s^2_{(\phi-\theta_{13})}s^2_{12}  -  s^2_{12}  \Delta m^2_{21}\nonumber  \\
 \lambda_-&=& a\, c^2_\phi ~~  +  ~~ \Delta m^2_{ee}\, s^2_{(\phi-\theta_{13})} .~ \nonumber  % \\[2mm]
%  \lambda_-&=& a\, c^2_\phi ~~  +  ~~ \Delta m^2_{31}\, s^2_{(\phi-\theta_{13})} ~~  + ~~ \Delta m^2_{21} \,c^2_{(\phi-\theta_{13})}s^2_{12}-  s^2_{12}  \Delta m^2_{21}  \nonumber  \\[2mm]
% \sum_{i=(1,\,2,\,3)} ~\lambda_{i}(dmp) & = & a+\Delta m^2_{31} +\Delta m^2_{21}  \nonumber  \\[2mm]
% \lambda_- (dmp)=\lambda_-+  s^2_{12}  \Delta m^2_{21} &=& a\, c^2_\phi ~~  +  ~~ \Delta m^2_{31}\, s^2_{(\phi-\theta_{13})} ~~  + ~~ \Delta m^2_{21} \,c^2_{(\phi-\theta_{13})}s^2_{12}  \nonumber  \\[2mm]
%  \sum_{i=(-,\,3)} ~\lambda_{i}(dmp) & = & a+\Delta m^2_{31} +\Delta m^2_{21} s^2_{12}  \nonumber  \\
%    \sum_{i=(1,\,2)} ~\lambda_{i}(dmp) & = & a\, c^2_\phi ~~+ ~~\Delta m^2_{31}\, s^2_{(\phi-\theta_{13})} ~~ + ~~\Delta m^2_{21} \,(c^2_{12} + c^2_{(\phi-\theta_{13})}s^2_{12})    \nonumber  
\end{eqnarray}
Adding $s^2_{12}  \Delta m^2_{21}$ to the eigenvalues, the convention used in  \cite{Denton:2016wmg}, simplifies $\lambda_{1}, \,\lambda_{2}$, but adds an additional term to  $\lambda_{3}, \,\lambda_{-}$.  Only the difference in the eigenvalues, $\Delta \lambda_{jk}\,$, are relevant for oscillations.

Eq. 6.1 is identical to a rewrite of eq. 2.4.5 of Denton et al. and eq. 5.1, 5.2 and 5.3 of Appendix II are identical to eq. 2.3.5, 2.4.9 and 2.4.5 of the same paper.

\section{Acknowledgements}

This manuscript has been authored by Fermi Research Alliance, LLC under Contract No. DE-AC02-07CH11359 with the U.S. Department of Energy, Office of Science, Office of High Energy Physics.

This project has received funding/support from the European Union's Horizon 2020 research and innovation programme under the Marie Sklodowska-Curie grant agreement No 690575.
 This project has received funding/support from the European Union's Horizon 2020 research and innovation programme under the Marie Sklodowska-Curie grant agreement No 674896.

HM is supported by Instituto F\'{\i}sica Te\'{o}rica, UAM/CSIC in Madrid, via ``Theoretical challenges of new high energy, astro and cosmo experimental data''  project, Ref: 201650E082. 

PBD acknowledges support from the Villum Foundation (Project No.~13164) and the Danish National Research Foundation (DNRF91 and Grant No.~1041811001).

\end{document}